\documentstyle[preprint,prl,aps,epsf]{revtex}
\begin{document}
\draft
\preprint{UTCCP-P-65, UTHEP-403}

\title{Quenched Light Hadron Spectrum}
\author{CP-PACS Collaboration\\
S.~Aoki$^1$, G.~Boyd$^2$, R.~Burkhalter$^{1,2}$, S.~Ejiri$^2$,
M.~Fukugita$^3$, S.~Hashimoto$^4$, Y.~Iwasaki$^{1,2}$, K.~Kanaya$^{1,2}$,
T.~Kaneko$^2$, Y.~Kuramashi$^4$, K.~Nagai$^2$,
M.~Okawa$^4$, H.~P.~Shanahan$^2$, A.~Ukawa$^{1,2}$, T.~Yoshi\'e$^{1,2}$}
\address{$^1$Institute of Physics, University of
Tsukuba, Tsukuba, Ibaraki 305-8571, Japan \\
$^2$Center for Computational Physics,
University of Tsukuba, Tsukuba, Ibaraki 305-8577, Japan \\
$^3$Institute for Cosmic Ray Research,
University of Tokyo, Tanashi, Tokyo 188-8502, Japan \\
$^4$High Energy Accelerator Research Organization
(KEK), Tsukuba, Ibaraki 305-0801, Japan}
\date{\today}
\maketitle

\tightenlines
\begin{abstract}
We present results of a large-scale simulation for the flavor non-singlet 
light hadron
spectrum in quenched lattice QCD with the Wilson quark action.
Hadron masses are calculated at four values of lattice spacing in the range 
$a \approx$ 0.1 -- 0.05 fm on lattices with a physical extent 
of 3 fm at five quark masses corresponding to 
$m_\pi/m_\rho \approx 0.75$ -- 0.4. 
The calculated spectrum in the continuum limit shows a systematic deviation
from experiment, though the magnitude of deviation is contained within 
11\%. 
Results for decay constants and light quark masses are also reported.
\end{abstract}
\pacs{11.15.Ha, 11.30.Rd, 12.15.Ff, 12.38.Gc, 14.20.-c, 14.40.-n}

Deriving the light hadron spectrum 
is a key step of lattice QCD in order to gain a convincing demonstration 
on the validity of QCD for strong interactions in the 
low-energy non-perturbative domain.
Under the present limitation of computer power, the first
possible step toward this goal is to establish the spectrum
in quenched QCD, in which the reaction of dynamical sea quarks is
switched off.  
The most extensive effort in this step, amongst numerous 
studies to date\cite{ref:review}, has been made by Butler 
{\it et al.}\cite{ref:GF11} using the Wilson quark action.  
Their results, extrapolated to the continuum limit and corrected for 
finite lattice size effects, show an 
agreement with the experimental spectrum within 6\% for 7 hadron masses 
calculated as compared to the estimated error of 1 to 9\% depending on 
particles.
A similar attempt with the Kogut-Susskind quark action\cite{ref:MILC-KS} 
has reported a nucleon mass higher than experiment by 3\%  with an error 
of 4\%.

In this article we report on our attempt at a quenched calculation 
with a precision significantly improved over those of previous studies, 
thereby establishing the quenched spectrum and simultaneously exploring the 
limitation of the quenched approximation.  This effort has been made 
possible by the CP-PACS computer, 
a massively parallel system with a peak speed of 614Gflops developed 
at the University of Tsukuba\cite{ref:cp-pacs}.
For preliminary reports of this work, see Ref.~\cite{ref:CP-PACS-preliminary}.

We carry out our study with the plaquette action for gluons and the
Wilson action for quarks.  Parameters employed in our simulations are
summarized in Table~\ref{tab:param}.
Four values of the coupling constant $\beta=6/g^2$ covering the 
range of lattice spacing $a\approx 0.1$ -- 0.05 fm are chosen, 
closer toward the continuum limit than 
$a\approx 0.14$ -- 0.07 fm explored in Ref.~\cite{ref:GF11}.
We employ lattices with a physical extent of $La \approx$ 3 fm
with which we expect negligible finite-size effects: 
no significant effect is observed for $La \geq$2 fm within
statistical error of about 2\%\cite{ref:MILC-W}.
We select five values of the hopping parameter $\kappa$ at which quark
mass takes values corresponding to $m_\pi/m_\rho\approx$ 0.75, 0.7, 0.6,
0.5 and 0.4, the last point being closer to the chiral limit than 
previous attempts with the Wilson action.
We abbreviate the two heavy quarks and the three light quarks as $s_i$ and 
$u_i$, respectively.

Gauge configurations are generated by the 5-hit pseudo-heat bath
algorithm with subsequent four over-relaxation updates. 
Quark propagators are solved on configurations fixed to the Coulomb gauge, 
using both the point source and an exponentially smeared source of 
a radius $r_0\approx 0.3 $ fm. 
For extraction of hadron masses, we use hadron propagators
constructed from only smeared sources, since they exhibit the earliest onset
of effective mass plateaus and the smallest statistical errors 
among various source combinations.
Masses are determined by correlated $\chi^2$ fits for degenerate hadrons 
as well as non-degenerate ones of the type $s_i u_j$ for mesons and 
$s_i s_i u_j$ and $s_i u_j u_j$ for baryons.

Chiral extrapolation is the most delicate issue in controlling the systematic 
errors.  For this extrapolation, most studies have employed low-order 
polynomials in quark masses.
Quenched chiral perturbation theory (Q$\chi$PT)
\cite{ref:QChPTSharpe,ref:QChPTBGmass,ref:QChPTBGdecay,ref:Booth,ref:LS},
on the other hand, predicts characteristic singularities in the chiral limit.

For pseudoscalar (PS) mesons made of quarks of masses $m_1$ and $m_2$, 
Q$\chi$PT mass formula reads\cite{ref:QChPTSharpe,ref:QChPTBGmass}
\begin{eqnarray}
 m_{PS,12}^2 &&= A(m_1+m_2)\{1 -\delta [\ln(2Am_1/\Lambda_\chi^2) \nonumber \\
         +&& m_2/(m_2-m_1) \ln(m_2/m_1) ] \} \nonumber \\ 
         +&& B(m_1+m_2)^2+O(m^3),
\label{eq:mps}
\end{eqnarray}
where the term proportional to $\delta$ represents the quenching effect. 
The following two quantities
\begin{eqnarray}
y &=& {\frac{2m_1}{m_1+m_2} \frac{m_{PS,12}^2}{m_{PS,11}^2}} \times 
   {\frac{2m_2} {m_1+m_2} \frac{m_{PS,12}^2}{m_{PS,22}^2}}, \\ \label{eq:x}
x &=& 2 + \frac{m_1+m_2}{m_1-m_2}\ln\Big(\frac{m_2}{m_1}\Big),\label{eq:y}
\end{eqnarray}
are then related by 
$y = 1 + \delta\! \cdot\! x+O(m^2)$.

\begin{figure}[bt]
\centerline{\epsfxsize=8.5cm \epsfbox{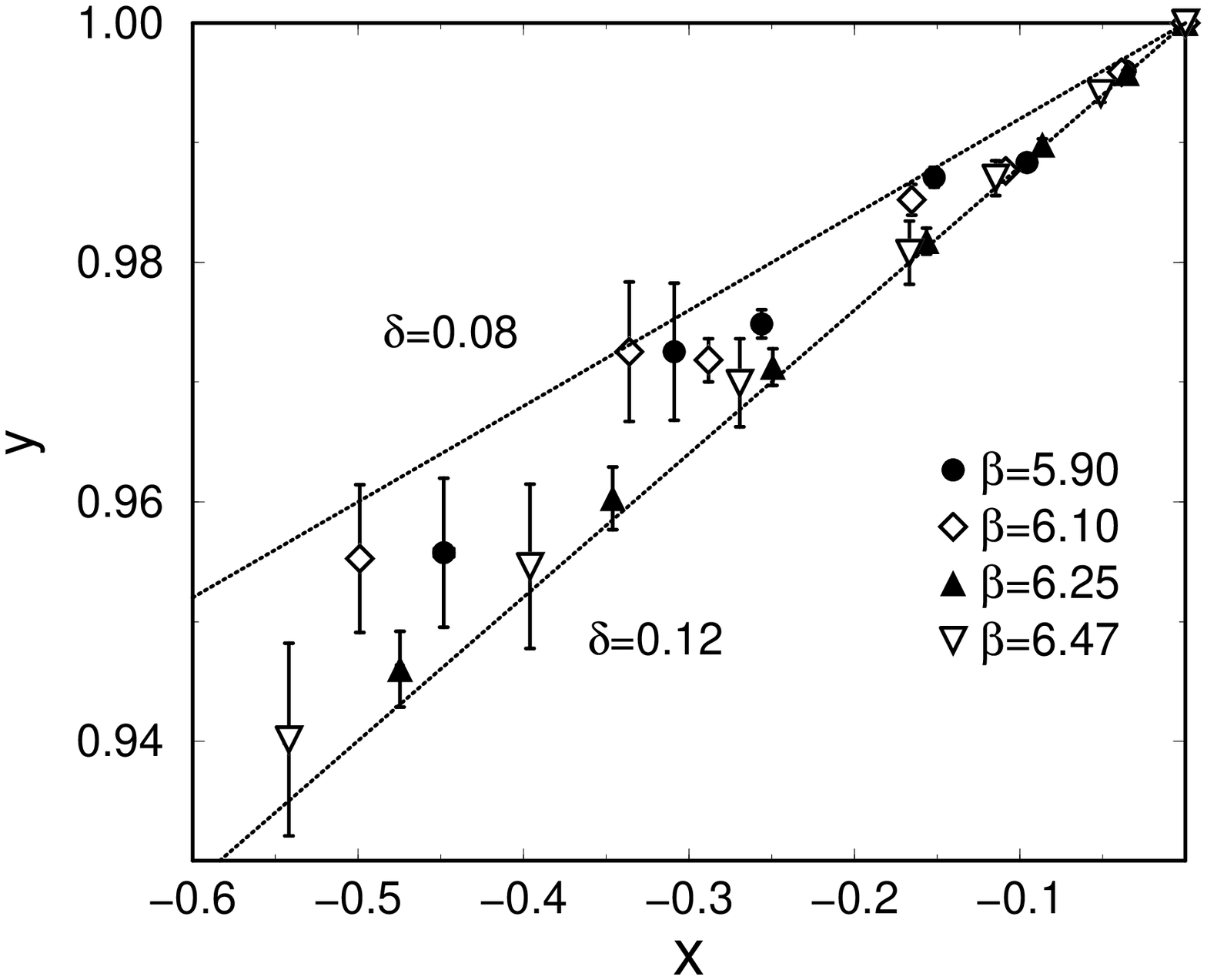}}
\caption{Results of a test for the presence of 
the quenched chiral logarithms in pseudoscalar meson mass data. 
See text for details.}
\label{fig:PSratio}
\end{figure}

Our PS meson mass data converted to $x$ and $y$ are shown in 
Fig.~\ref{fig:PSratio}.  
The quark mass determined from the axial Ward 
identity\cite{ref:Bo,ref:Itoh86} 
($m_q^{AWI}$) is used since it has neither quenched chiral
logarithms\cite{ref:GS} nor ambiguity due to chiral extrapolations.
We observe a clear non-zero slope in the range $\delta \approx 0.08$ -- 0.12. 
The ratio $m_{PS,12}^2/m_{PS,11}^2$ used in Ref.~\cite{ref:QChPTBGmass}
receives a correction from the $O(m^2)$ term in (\ref{eq:mps}), and 
the plot does not fall on a common line. 

Similarly, an examination of the relation $y=1- (\delta/2) \cdot x$
expected for the 
decay constant ratio $y=f_{12}^2/(f_{11} f_{22})$\cite{ref:QChPTBGdecay} 
gives a non-zero value of $\delta \approx$ 0.08 -- 0.16.

Finally, we make a correlated fit of the PS mass results to the Q$\chi$PT mass 
formula.  For this fit the vector Ward identity quark mass 
$m_q^{VWI} = (1/\kappa-1/\kappa_c)/2$ is used, taking $\kappa_c$ as 
one of the fit parameters. The fitting, made independently for
degenerate and non-degenerate cases, yields 
mutually consistent results for $\delta$  
in the range $\delta \approx 0.06$ -- 0.12 
when $\Lambda_\chi$ is varied over $\Lambda_\chi \approx 0.6$ -- 1.4 GeV.

A noticeable result from this fit is that the value of $\kappa_c$ 
agrees well with $\kappa_c^{AWI}$ at which linearly
extrapolated $m_q^{AWI}$ vanishes with at most 2.8$\sigma$ discrepancy, whereas
quadratic (cubic) extrapolation of $m_{PS}^2$ in $m_q^{VWI}$ results in
$\kappa_c$ differing as much as 17$\sigma$ (12$\sigma$).
This observation suggests that the Q$\chi$PT fit is legitimate
as $m_{PS}^2$ and $m_q^{AWI}$ should vanish at the same $\kappa$ value
from their definitions.

We take these results as strong support that our PS meson results 
are consistent with the presence 
of quenched chiral logarithms with a magnitude $\delta\approx 0.1$. 
We should mention that there is in (\ref{eq:mps}) another quenched 
singularity 
term of the form $\alpha_\Phi m_i^2\ln m_i$\cite{ref:QChPTBGmass}.  
Our mass data do not show clear indications for the presence of this term. 
  
Q$\chi$PT mass formulae for vector mesons\cite{ref:Booth} and
baryons\cite{ref:LS} can be schematically written as
\begin{equation}
m_H(m_{PS}) = m_0 + C_{1/2} m_{PS} + C_1 m_{PS}^2 + C_{3/2}m_{PS}^3,
\label{eq:mvb}
\end{equation}
where $C_i$ are polynomials of the couplings of the 
quenched chiral Lagrangian.  The coefficient $C_{1/2}$, which is proportional 
to $\delta$, represents the quenched singular term.
Encouraged by the results for the PS mesons, we attempt to fit our 
mass data to (\ref{eq:mvb}).  Fully constraining the large number of coupling 
parameters (6 for vector mesons and 11 for baryons 
in addition to $\delta$ and $\alpha_\Phi$) is difficult
under the limitation of the accuracy of our mass data.
We therefore
employ $\delta=0.1$ and $\alpha_\Phi=0$, 
and drop the couplings of the flavor-singlet PS 
meson to vector mesons and baryons. 
Fits are made to degenerate and non-degenerate data together ignoring 
correlations, as we find correlated fits to be quite unstable. 
Errors are calculated with the single elimination jackknife method.

\begin{figure}[bt]
\centerline{\epsfxsize=8.5cm \epsfbox{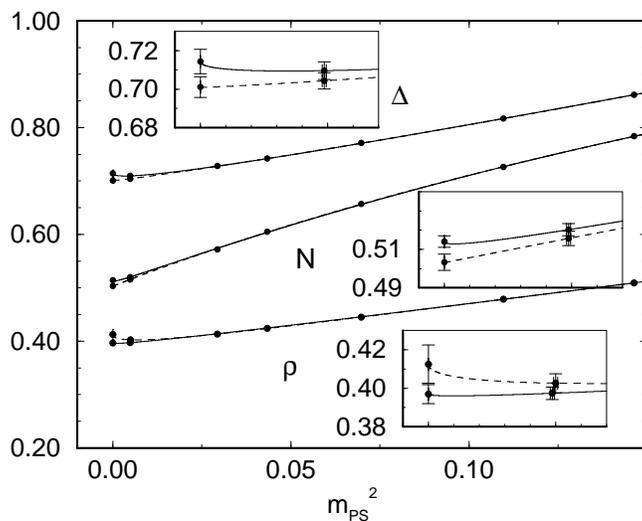}}
\caption{Degenerate hadron masses vs. $m_{PS}^2$ at $\beta$=5.9.
The leftmost points are values extrapolated to the chiral limit, 
and the second ones from left are those at the physical point. 
Fitting curves from two types of chiral extrapolations based on 
Q$\chi$PT are reproduced. See text for details.}
\label{fig:chiral}
\end{figure}

With this procedure we obtain a good fit with small $\chi^2/{\rm dof} < 0.8$ 
keeping all $O(m_{PS})$ and $O(m_{PS}^2)$ terms for vector mesons and 
decuplet baryons.
For octet baryons, we also include $O(m_{PS}^3)$ terms since
the nucleon mass shows a clear negative curvature, 
which is opposite to the predicted negative sign of $C_{1/2}$ 
(see Fig.~\ref{fig:chiral}).
We drop the octet-decuplet coupling because the decuplet baryon fits
lead to a value of this coupling consistent with zero.
These fits yield $C_{1/2}$ = $-0.071(8)$ for $\rho$, $-0.118(4)$ for nucleon
and $-0.14(1)$ for $\Delta$ when averaged over the four $\beta$ 
values of the simulation.  These values are much smaller than phenomenological
estimates: $C_{1/2}= -0.71$ for $\rho$\cite{ref:Booth} 
and $-0.27$ for nucleon\cite{ref:LS}. 

We have also attempted to include the $O(m_{PS}^3)$ terms 
in the fit for vector 
mesons and decuplet baryons, and the octet-decuplet coupling for octet 
baryons, since it is theoretically more consistent to include them.  
Making this more general fit, we find the values of couplings to change 
significantly. In particular the coefficient
of the $O(m_{PS})$ term becomes $C_{1/2} \approx $ $-0.29$ for $\rho$,
$-0.01$ for nucleon and $-0.02$ for $\Delta$.
The stability of fit worsens considerably, however, compared to the fit
above: the covariance matrix of the fit is close to singular, 
and some of the couplings exhibit significant variation with the 
lattice spacing.
Nonetheless, as we illustrate in Fig.~\ref{fig:chiral} for 
representative hadron masses,
the curves from the two fits (dashed lines for the general fit and solid 
lines for the simplified fit as described in the previous paragraph) 
are indistinguishable in the range of measured points.  
The deviation remains small even at the physical point (second points from 
left in Fig.~\ref{fig:chiral}).
 
We conclude that
the accuracy of the mass results and the covered range of quark mass
are not sufficient to establish the presence of the quenched singular term
for vector mesons and baryons.
We adopt the former fit with simplified Q$\chi$PT formulae
in the subsequent analyses as it is overall
more stable.  

We emphasize that adopting the latter general fit does not affect
our conclusions below, as changes in hadron masses at the physical point 
are 5\% ($5\sigma$) or less at finite lattice spacings and at most 
1.2\% ($1.3\sigma$) after the continuum extrapolation. 
We have also checked that varying $\delta$ over $\delta\approx$ 
0.08 -- 0.12 ($\alpha_\Phi$ over $\alpha_\Phi \approx$
$-0.7$ -- $+0.7$) changes the mass results by 
no more than 0.4\% (1.3$\sigma$) (2.9\%(4.7$\sigma$))
at finite lattice spacings and 
0.3\% (0.3$\sigma$) (2.2\%(1.4$\sigma$)) in the continuum limit.

\begin{figure}[bt]
\centerline{\epsfxsize=8.5cm \epsfbox{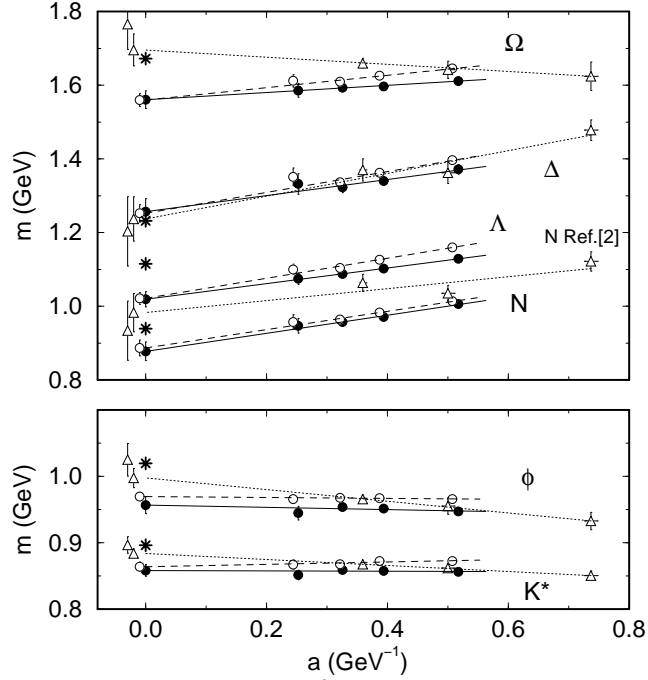}}
\caption{Typical continuum extrapolations of masses with $m_K$ as input.
Filled (open) circles are data from the Q$\chi$PT (polynomial) fits.
Open triangles are from Butler {\it et al.}\protect\cite{ref:GF11},
the leftmost ones being their final results after finite size corrections. 
Stars represent experimental values.}
\label{fig:continuum}
\end{figure}

The physical point for up and down quarks and
the lattice scale are fixed from the experimental values of 
$m_\pi$ = 0.1350 GeV and $m_\rho$ = 0.7684 GeV, 
and the strange quark mass by that of $m_K$ = 0.4977 GeV or $m_\phi$ = 1.0194 GeV.  
We then extrapolate the results linearly in the lattice spacing to the continuum
limit. 
We find that $\chi^2/{\rm dof}$ of the fit with only
the leading scaling violation term is reasonably small ($< 1.6$)
for each hadron. The statistical error in the continuum limit is 1--3\%.   
In Fig.~\ref{fig:continuum} are shown typical continuum extrapolations.
The magnitude of scaling violation at $a=0.075$~fm, the mid-value of
the range of our lattice spacing, is at most 10\%. 
We then expect that $O(a^2)$ terms have an effect of 1\% or less,
which is smaller than the statistical error.

In order to examine how much our results in the continuum limit depend
on the Q$\chi$PT fit, we repeat the analysis with 
the conventional polynomial fit of hadron masses in terms of $1/\kappa$. 
This fit (cubic for nucleon, quadratic for
others including PS mesons) yields masses at the physical point differing 
by at most 3\% at finite lattice spacings relative to the Q$\chi$PT fit,
and less than 1.6\% (1.6$\sigma$) in the continuum limit. 
The continuum extrapolations for the Q$\chi$PT and polynomial chiral 
fits are compared in Fig.~\ref{fig:continuum}, with 
filled circles and solid lines for the former and open circles and 
dashed lines for the latter.

\begin{figure}[bt]
\centerline{\epsfxsize=8.5cm \epsfbox{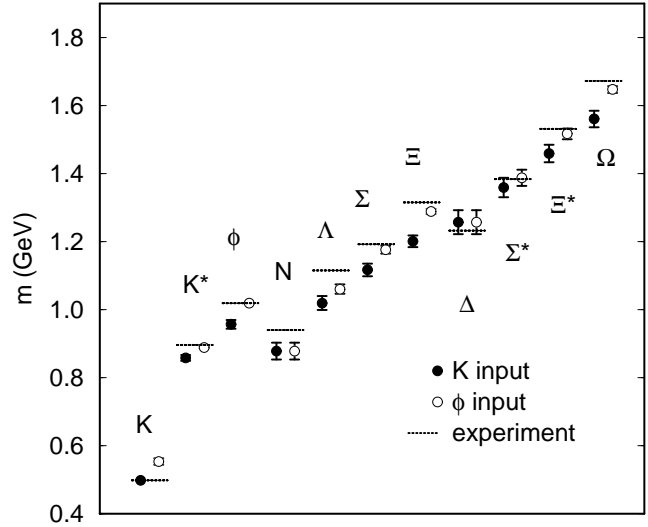}}
\caption{Final spectrum results compared to experiment.}
\label{fig:spectrum}
\end{figure}

We present our final result for the quenched light hadron spectrum 
in Fig.~\ref{fig:spectrum}.  
The numerical values are given in Table~\ref{tab:spectrum}.
Systematic errors, being comparable or less than the statistical ones, 
are not included.

Our chief finding, quite clear from Fig.~\ref{fig:spectrum}, 
is that the quenched spectrum systematically deviates from the experimental 
spectrum. 
Quantitatively stated, if one uses the $K$ meson mass to fix the strange 
quark mass, the vector meson masses $m_{K^*}$ and $m_\phi$
are smaller by 4\% (4$\sigma$) and 6\% (5$\sigma$), respectively, 
the octet baryon masses are smaller by 6--9\% (4--7$\sigma$), 
and so are the decuplet mass splittings (30\% on average). 
Alternatively, if $m_\phi$ is employed to fix the strange quark mass, 
$m_{K^*}$ agrees with experiment within 0.8\% (2$\sigma$) and
the discrepancies for baryon masses are much reduced.  However, 
$m_K$ is larger by 11\% (6$\sigma$).  
As one sees in Table~\ref{tab:spectrum} this
11\% represents the largest deviation between our results and the 
experiment.

Our finding of a small mass splitting between $K$ and $K^*$ 
differs from that of Butler {\it et al.}\cite{ref:GF11} 
who found an agreement with experiment.
Another difference is the masses of $\Xi^*$ and 
$\Omega$ baryon determined with $m_K$ as input, 
for which they reported values higher than experiment by 
3--5\%, while our results are smaller by a similar magnitude. 
How these differences arise are shown in Fig.~\ref{fig:continuum} where 
the results of Ref.~\cite{ref:GF11} are plotted by open triangles. 
For the nucleon both the results from Butler {\it et al.}\cite{ref:GF11} and
the MILC collaboration\cite{ref:MILC-KS} are consistent with experiment;
our value is smaller by 7\% (2.5$\sigma$). 

We also calculate PS decay constants and quark masses
using tadpole-improved one-loop values for renormalization constants.
For the PS decay constant we find 
$f_\pi=120.0(5.7)$~MeV and $f_K=138.8(4.4)$~MeV with $m_K$ as input,
which are smaller than experiment by 9\%(2$\sigma$) and 13\%(5$\sigma$),
respectively.  
Quark masses are determined by a combined linear continuum extrapolation of
$m_q^{VWI}$ and $m_q^{AWI}$, since the large difference of values from the
two definitions at finite lattice spacings\cite{ref:Itoh86,ref:LANL} 
vanishes toward the continuum limit\cite{ref:CP-PACS-preliminary}. 
We obtain $m_{u,d}$=4.55(18) MeV, and $m_s$=115.1(2.3) MeV ($m_K$ input) or 
142.9(5.8) MeV ($m_\phi$ input) in the $\overline{\rm MS}$ scheme 
at $\mu=2$~GeV.
A 20\% disagreement between the two values for $m_s$ originates from 
the small meson hyperfine splitting, and hence
represents a quenching effect.

In conclusion we have found that the light hadron spectrum in quenched 
QCD systematically deviates from the experimental spectrum when examined 
with accuracy better than a 10\% level.
In the course of analyses 
we have observed a strong support for the presence of quenched chiral 
singularities for pseudoscalar mesons.  Whether vector mesons and 
baryons also have such singularities, however, remains as a problem for 
future investigations.

We thank all the members of the CP-PACS Project with whom the
CP-PACS computer has been developed. 
Valuable discussions with M.~Golterman and S.~Sharpe are gratefully 
acknowledged. 
This work is supported in part by the Grants-in-Aid of Ministry of
Education (Nos. 08NP0101 and 09304029).
GB, SE, and KN are JSPS Research Fellows. 
HPS is supported by JSPS Research for Future Program.

\begin{table}[b]
\caption{Simulation parameters.}
\label{tab:param}
\begin{center}
\begin{tabular}{lccccc}
$\beta$     &  $L^3 T$  & $a^{-1}$(GeV) & $La$(fm) & \#conf. &
sweep/conf  \\ 
\hline
5.90 & $32^3\,\, 56$  & 1.934(16) & 3.26(3) & 800 & 200 \\
6.10 & $40^3\,\, 70$  & 2.540(22) & 3.10(3) & 600 & 400 \\
6.25 & $48^3\,\, 84$  & 3.071(34) & 3.08(3) & 420 & 1000 \\
6.47 & $64^3 112$     & 3.961(79) & 3.18(6) & 150 & 2000 \\
\end{tabular}
\end{center}
\end{table}

\begin{table}[t]
\caption{
Spectrum results. Deviation from experiment with its statistical significance
is also given.}
\label{tab:spectrum}
\begin{center}
\begin{tabular}{ll|lr|lr}
\         &          & 
\multicolumn{2}{c|}{$m_K$ input} &
\multicolumn{2}{c}{$m_\phi$ input} \\
\         & exp.  & mass(GeV) & deviation & mass(GeV) & deviation\\
\hline 
$K$       & 0.4977 &          &      & 0.553(10) & 11.2\% 5.6$\sigma$\\
$K^*$     & 0.8961 & 0.858(09) & $-$4.2\% 4.3$\sigma$ & 
                     0.889(03) & $-$0.8\% 2.3$\sigma$ \\
$\phi$    & 1.0194 & 0.957(13) & $-$6.1\% 4.8$\sigma$ & 
                               &                      \\
$N$       & 0.9396 & 0.878(25) & $-$6.6\% 2.5$\sigma$ &
                     0.878(25) & $-$6.6\% 2.5$\sigma$ \\
$\Lambda$ & 1.1157 & 1.019(20) & $-$8.6\% 4.7$\sigma$ &
                     1.060(13) & $-$5.0\% 4.1$\sigma$ \\
$\Sigma$  & 1.1926 & 1.117(19) & $-$6.4\% 4.1$\sigma$ &
                     1.176(11) & $-$1.4\% 1.5$\sigma$ \\
$\Xi$     & 1.3149 & 1.201(17) & $-$8.7\% 6.8$\sigma$ &
                     1.288(08) & $-$2.0\% 3.5$\sigma$ \\
$\Delta$  & 1.2320 & 1.257(35) &    2.0\% 0.7$\sigma$ &
                     1.257(35) &    2.0\% 0.7$\sigma$ \\
$\Sigma^*$& 1.3837 & 1.359(29) & $-$1.8\% 0.9$\sigma$ &
                     1.388(24) &    0.3\% 0.2$\sigma$ \\
$\Xi^*$   & 1.5318 & 1.459(26) & $-$4.7\% 2.8$\sigma$ &
                     1.517(16) & $-$1.0\% 0.9$\sigma$ \\
$\Omega$  & 1.6725 & 1.561(24) & $-$6.7\% 4.7$\sigma$ &
                     1.647(10) & $-$1.5\% 2.6$\sigma$ \\
\end{tabular}
\end{center}
\end{table}


\begin{references}
\bibitem{ref:review} 
For recent reviews see, 
R.~Kenway, hep-lat/9810054;\ 
T.~Yoshi\'e, Nucl. Phys. {\bf B} (Proc. Suppl.) {\bf 63}, 3 (1998). 

\bibitem{ref:GF11} 
F.~Butler {\it et al.}, Nucl. Phys. {\bf B430}, 179 (1994).

\bibitem{ref:MILC-KS}
MILC Collaboration, C.~Bernard {\it et al.},
Phys. Rev. Lett. {\bf 81}, 3087 (1998).

\bibitem{ref:cp-pacs} 
Y.~Iwasaki, Nucl. Phys. {\bf B} (Proc. Suppl.) {\bf 53}, 1007(1997);
T.~Boku, K.~Itakura, H.~Nakamura, K.~Nakazawa, in {\it Proceedings of 
ACM International Conference on Supercomputing'97}, 108 (1997). 

\bibitem{ref:CP-PACS-preliminary} 
CP-PACS Collaboration, S.~Aoki {\it et al.}, 
Nucl. Phys. {\bf B} (Proc. Suppl.) {\bf 60A}, 14 (1998);
{\it ibid.} {\bf 63}, 161 (1998); 
hep-lat/9809146; R.~Burkhalter, hep-lat/9810043. 

\bibitem{ref:MILC-W} 
MILC Collaboration, C.~Bernard {\it et al.}, 
Nucl. Phys. {\bf B} (Proc. Suppl.) {\bf 60A}, 3 (1998).

\bibitem{ref:QChPTSharpe}
S.~R.~Sharpe, Phys. Rev. D {\bf 46}, 3146 (1992).

\bibitem{ref:QChPTBGmass}
C.~W.~Bernard and M.~F.~L.~Golterman, Phys. Rev. D {\bf 46}, 853 (1992).

\bibitem{ref:QChPTBGdecay}
C.~W.~Bernard and M.~F.~L.~Golterman, 
Nucl. Phys. {\bf B} (Proc. Suppl.) {\bf 30}, 217 (1993).

\bibitem{ref:Booth}
M.~Booth {\it et al.}, Phys. Rev. D {\bf 55}, 3092 (1997).

\bibitem{ref:LS}
J.~N.~Labrenz and S.~R.~Sharpe, Phys. Rev. D {\bf 54}, 4595 (1996).

\bibitem{ref:Bo}
M.~Bochicchio {\it et al.}, Nucl. Phys. {\bf B262}, 331 (1985).

\bibitem{ref:Itoh86}
S.~Itoh {\it et al.},
Nucl. Phys. {\bf B274}, 33 (1986).

\bibitem{ref:GS} M.~Golterman and S.~Sharpe, private communications.

\bibitem{ref:LANL}
T.~Bhattacharya, R.~Gupta, G.~Kilcup and S.~Sharpe,
Phys. Rev. D {\bf 53}, 6486 (1996).

\end{references}
\end{document}